\documentclass[prb,twocolumn,aps,superscriptaddress,amsmath,amssymb]{revtex4}
\usepackage{dcolumn}

\usepackage{amsmath,amssymb,graphicx,bm,xcolor}

\begin{document}

\title{How Infrared Singularities Affect Nambu-Goldstone Bosons
\\
at Finite Temperatures}

\author{Takafumi K{\sc ita}}
\affiliation{Department of Physics, Hokkaido University, Sapporo 060-0810, Japan}

\begin{abstract}
Ordered phases realized through broken continuous symmetries embrace long-range order-parameter fluctuations as manifest in the power-law decays 
of both the transverse and longitudinal correlations, which are similar to those at the second-order transition point.
We calculate the transverse one-loop correction to the dispersion relation of nonrelativistic Nambu-Goldstone (NG) bosons at finite temperatures,
assuming that they have well-defined dispersion relations with some integer exponents.
It is found that the transverse correlations make the lifetime of the NG bosons vanish right on the assumed dispersion curve below three dimensions at finite temperatures. 
Combined with the vanishing of the longitudinal ``mass'', the result indicates 
that the correlations generally  bring both an intrinsic lifetime to the excitations
and a qualitative change in the dispersion relation at long wavelengths
below three dimensions at finite temperatures, unless it is protected by some conservation law.
A more detailed calculation performed specifically on a single-component Bose-Einstein condensate reveals 
that the gapless Bogoliubov mode predicted by the perturbative treatment
is modified substantially after incorporating the correlations to have (i) an intrinsic lifetime and 
(ii) a peak shift  at long wavelengths from the linear dispersion relation.
Thus, the NG bosons may be fluctuating intrinsically into and out of the ordered ``vacuum'' 
to maintain the order temporally.
\end{abstract}

\maketitle

\section{Introduction}

Goldstone's theorem \cite{Nambu61,Goldstone61,GSW62,Weinberg95} states that
broken continuous symmetries necessarily accompany gapless excitations called Nambu-Goldstone (NG) bosons in the ordered phases.
Examples include spin waves of isotropic ferromagnets \cite{Bloch30,HP40,Dyson56} and antiferromagnets,\cite{Anderson52,Kubo52} and the Bogoliubov mode of single-component Bose-Einstein condensates (BECs).\cite{Bogoliubov47,LHY57,Beliaev58,HP59,GN64}
Their existence has been elucidated theoretically prier to the proof of the theorem \cite{GSW62} based on perturbative treatments around zero temperature.\cite{Bloch30,HP40,Anderson52,Kubo52,Dyson56,Bogoliubov47,LHY57,Beliaev58,HP59,GN64}
The connection among the number of broken continuous symmetries, that of emergent NG bosons, and their dispersion relations has been clarified
for non-relativistic systems \cite{NC76,Schafer01,WB11,WM12,Hidaka13}
by implicitly assuming existence of well-defined (long-lived) excitations and analyticity in their dispersion relations. 

On the other hand, the gapless excitations have been known to 
cause infrared-singular behaviors in the ordered phases similar to those at the second-order transition point.\cite{Ma76,Amit84,Justin96}
Most notably,  the transverse order-parameter correlation function diverges as $k^{-2}$ in the wave vector space for ${\bm k}\!\rightarrow\! {\bm 0}$,
as can be shown rigorously  based on the Bogoliubov inequality,\cite{Hohenberg67,MW66} to make any long-range order impossible below {\it two} dimensions
at finite temperatures; the exponent $-2$ without the anomalous dimension is crucial in reaching the conclusion.\cite{Hohenberg67}
Moreover, the longitudinal correlation function also diverges as $k^{d-4}$ in $d$ dimensions for $2\!\leq d\!<\!4$ due to the transverse correlations, as can be concluded based on various techniques.\cite{ABDS99}
Hence, the longitudinal order-parameter susceptibility, 
which is predicted to recover a finite value in the ordered phases by the mean-field analysis,\cite{Ma76,Amit84}
remains divergent from the second-order transition point down to zero temperature.\cite{Amit84,ABDS99}
Indeed, this fact has been found separately and independently for specific systems 
such as isotropic ferromagnets \cite{HP40,Larkin67,SM70,BWW73,PP73} 
and single-component BECs.\cite{NN75}
Thus,  the transverse correlations bring a qualitative change 
in those of the longitudinal branch from {\it massive} to {\it massless} in the terminology of relativistic quantum field theory.\cite{Weinberg95,ABDS99}

The purpose of the present paper is to 
study how the infrared singularities affect the NG bosons themselves,
which have been regarded as well-defined quasiparticles in the transverse channel
having definite dispersion relations with some integer exponents.\cite{NC76,Schafer01,WB11,WM12,Hidaka13}
For example, the Bogoliubov theory on the single-component BECs with a single broken symmetry in the phase angle  
predicts that there should be a gapless excitation with a linear dispersion relation at long wavelengths,\cite{Bogoliubov47,OKSD05} 
in agreement with the general argument of assuming analyticity.\cite{WM12}
However, the predicted linearity results from the assumption inherent in the perturbative treatment that
the longitudinal susceptibility is finite, which no longer holds after the correlations through the transverse channel are incorporated, 
as mentioned above. 
Hence, one may expect a substantial modification of the long-wavelength dispersion relation from the linear one
with an infinite lifetime.

It should be noted that the topic on the BECs has already been studied at zero temperature \cite{NN75,CCPS97} to conclude that
the Bogoliubov mode no longer exists because of the infrared singularities to be replaced by another gapless branch with a linear dispersion relation.
An implicit and crucial assumption in reaching the result is again the analyticity of the inverse Green's function $\hat{G}^{-1}(\vec{k})$
at $\vec{k}\!=\!\vec{0}$,
on which it is  expanded in $\vec{k}\!=\!({\bm k},\varepsilon)$ up to the second order, 
especially up to $\varepsilon^2$ in terms of the frequency $\varepsilon$ beyond the standard linear order
to find the new mode.
However, the presumed analyticity is not self-evident at all in view of the divergence of the longitudinal susceptibility; hence, existence of the distinct phonon-like mode at $T\!=\!0$ in the single-particle branch remains unestablished yet.
This is more so at finite temperatures where 
the Bose distribution function $f(\varepsilon)$ with a pole at $\varepsilon\!=\!0$ is relevant,
so that the singular $\varepsilon\!=\!0$
branch should be treated separately from those of $\varepsilon\!\neq\! 0$,
similarly as in the treatment of the critical phenomena.\cite{Ma76,Amit84,Justin96}

This paper is organized as follows. In Sec.\ 2, we give a general estimate on the lifetime of a Nambu-Goldstone boson
assuming a well-defined dispersion relation initially and incorporating the transverse correlations perturbatively.
In Sec.\ 3, we present a more detailed calculation on the excitation spectrum of a single-component Bose-Einstein condensate.
In Sec.\ 4, we provide concluding remarks. We use the units of $\hbar=k_{\rm B}=1$.

\section{Analytic Estimation of Lifetime}

In this section, we present an analytic one-loop estimation of the lifetime of an NG boson at finite temperatures,
which is assumed to have a well-defined dispersion relation $E_k \propto k^n$ with $n\geq 1$.
Unlike previous studies,\cite{HM65,Pitaevskii97,Giorgini98} we carry it out by retaining the full spectral resolution 
in terms of the wave vector; see the last paragraph of this section on this point.

The corresponding transverse Green's function $G_{t}^{}$ can be expressed in the Lehmann representation\cite{AGD63,FW71}
as
\begin{align}
G_t^{}(k,z)\!=\! \int_{-\infty}^\infty \frac{d\varepsilon}{2\pi} \frac{\rho_t^{}(k,\varepsilon)}{z-\varepsilon}.
\label{G_t}
\end{align}
where $\rho_t^{}$ is the spectral function
\begin{align}
\rho_t^{}(k,\varepsilon)=2\pi [w_k^+\delta(\varepsilon-E_k)- w_k^-\delta(\varepsilon+E_k)],
\label{rho_t-QP}
\end{align}
given in terms of the excitation energy $E_k>0$ and its spectral weight $w_k^\pm \geq 0$.
The weight $w_k^- $ becomes finite when the relevant broken symmetry yields a connection between the particle and hole channels,
as in the cases of BEC and antiferromagnetism.
It has been shown generally\cite{Hohenberg67,MW66} that $G_t^{}(k,z=0)\propto k^{-2}$ holds in the limit of $k\rightarrow 0$,
which for Eq.\ (\ref{G_t}) with Eq.\ (\ref{rho_t-QP}) can be expressed alternatively as
\begin{align}
\frac{w_k^{\pm}}{E_k} \propto k^{-2} .
\label{w_k/E_k}
\end{align}

We now consider the transverse one-loop process that is responsible for
the infrared singularities of the continuous-symmetry-breaking phases.
It can be expressed analytically in terms of $G_t^{}(k,z)$ by 
\begin{align}
\chi_{tt}^{}(q,i\omega_\ell)\equiv -\frac{T}{V}\!\sum_{\bm k}\!\sum_n G_t^{}(|{\bm k}+{\bm q}|,i\varepsilon_n+i\omega_\ell)G_t^{}(k,i\varepsilon_n),
\label{chi_tt}
\end{align}
where $T$ and $V$ are the temperature and  volume, respectively, 
and $\varepsilon_n\!\equiv\! 2n\pi T$ and $\omega_\ell \!\equiv\! 2\ell\pi T$ ($n,\ell\!=\!0,\pm 1,\cdots$) are boson Matsubara frequencies $(\hbar\!=\!k_{\rm B}\!=\! 1)$.
Substituting Eq.\ (\ref{G_t}) and using the Bose distribution function $f(\varepsilon)\equiv (e^{\varepsilon/T}-1)^{-1}$, 
we can transform the sum over $n$ in Eq.\ (\ref{chi_tt}) into a contour integral, collect residues on the real axis, and perform the analytic continuation $i\omega_\ell\rightarrow \omega+i0_+$.\cite{AGD63,FW71}
We thereby obtain the imaginary part of the retarded response function as
\begin{align}
{\rm Im}\chi_{tt}^{\rm R}(q,\omega)=&\,\frac{1}{2V}\sum_{\bm k}\int_{-\infty}^\infty \frac{d\varepsilon}{2\pi}
\rho_t^{}(k,\varepsilon)\rho_t^{}(|{\bm k}+{\bm q}|,\varepsilon+\omega)
\notag \\
&\,\times [f(\varepsilon+\omega)- f(\varepsilon)] .
\label{ImChi_tt}
\end{align}
Substitution of Eq.\ (\ref{rho_t-QP}) into Eq.\ (\ref{ImChi_tt}) yields
\begin{align}
&\,{\rm Im}\chi_{tt}^{\rm R}(q,\omega)
\notag \\
=&\,-\frac{\pi}{V}\sum_{\bm k}\sum_{\sigma=\pm} \biggl\{f(E_k)
w_k^\sigma w_{|{\bm k}+{\bm q}|}^\sigma \delta(\omega-E_{|{\bm k}+{\bm q}|}+E_k)
\notag \\
&\,+\left[f(E_k)+\frac{1}{2}\right]
w_k^\sigma w_{|{\bm k}+{\bm q}|}^{-\sigma} \delta(\omega-E_{|{\bm k}+{\bm q}|}-E_k)
\notag \\
&\,\,-\,(\omega\rightarrow -\omega)\biggr\},
\label{ImChi_tt-2}
\end{align}
where we have transformed $f(-E)\!=\!-1-f(E)$ for $E>0$ and subsequently made a change of variables ${\bm k}+{\bm q}\!=\!-{\bm k}'$ 
for terms with the factor $f(E_{|{\bm k}+{\bm q}|})$ or $f(E_{|{\bm k}+{\bm q}|})+\frac{1}{2}$.

Let us focus on the first term in the curly brackets of Eq.\ (\ref{ImChi_tt-2}), where we can estimate
\begin{subequations}
\label{fac1}
\begin{align}
f(E_k)w_k^\sigma\propto (T/E_k)w_k^\sigma \propto k^{-2} .
\label{fac11}
\end{align}
 for $k\rightarrow 0$ at $T>0$ based on Eq.\ (\ref{w_k/E_k}).
Note that it is the pole $\varepsilon\!=\!0$ of $f(\varepsilon)$ that brings the singularity.
Moreover, integration of $w_{|{\bm k}+{\bm q}|}^{\sigma}\delta(\omega-E_{|{\bm k}+{\bm q}|}+E_k)$ over the angle $\theta\equiv \arccos({\bm k}\cdot{\bm q}/kq)$
yields another factor $k^{-1}$ at $\omega=E_q$,
\begin{align}
&\,\int_0^\pi d\theta\sin^{d-2}\theta \,w_{|{\bm k}+{\bm q}|}^\sigma\delta(\omega-E_{|{\bm k}+{\bm q}|}+E_k)
\notag \\
=&\,\frac{1}{kq} \int_{E_{|k-q|}}^{E_{k+q}} dE_{k_1} \frac{k_1 \sin^{d-3}\theta}{dE_{k_1}/dk_1} w_{k_1}^\sigma \delta(\omega-E_{k_1}+E_k)
\notag \\
\stackrel{\omega=E_q}{\longrightarrow} &\,\frac{1}{kq} \left[\frac{k_1 \sin^{d-3}\theta}{dE_{k_1}/dk_1} w_{k_1}^\sigma \right]_{E_{k_1}=E_q+E_k } 
\propto \frac{1}{k} ,
\label{fac12}
\end{align}
\end{subequations}
where we have made a successive change of variables $\theta\rightarrow k_1\equiv |{\bm k}+{\bm q}| \rightarrow E_{k_1}$.
Since $E_{|k-q|}< E_q+E_k\leq E_{k+q}$ holds for $E_k\propto k^n$ with $n\geq 1$,
the factor in the square brackets in the third line is finite and positive; see the original expression of the first line on the latter point.

The same transformation can be performed for the other terms with $f(E_k)$ in the curly brackets of Eq.\ (\ref{ImChi_tt-2}). 
They yield additional contributions for which $E_{k_1}\!=\!E_q+E_k$ in the third line of Eq.\ (\ref{fac12}) 
is replaced by $E_{k_1}\!=\!E_q-E_k$, $E_{k_1}\!=\!-E_q+E_k$, and $E_{k_1}\!=\!-E_q-E_k$, respectively.
However, the latter two conditions cannot be satisfied in the limit of $k\rightarrow 0$, as seen from the fact that they reduce to $E_q=-E_q$.
Hence, the latter two terms vanish for $k\rightarrow 0$,
leaving the first two finite contributions that are additive.

\begin{figure}[t]
\begin{center}
\includegraphics[width=0.9\linewidth]{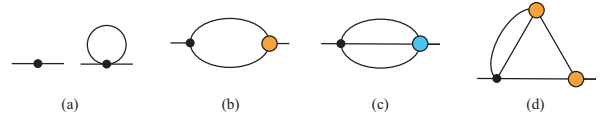}
\end{center}
\caption{(Color online) Self-energy diagrams. Small and medium circles represent symmetrized bare and renormalized vertices, respectively.}
\label{Fig1}
\end{figure}

It follows from Eq.\ (\ref{fac1}) that the integrand of Eq.\ (\ref{ImChi_tt-2}) at $\omega=E_q$ with $q>0$ behaves as $k^{(d-1)-3}$ for $k\rightarrow 0$.
Hence, ${\rm Im}\chi_{tt}^{\rm R}(q,\omega)$ at $\omega\!=\! E_q$ is concluded to diverge for $d\leq 3$ at finite temperatures.
We also note that, with the emergence of three-point vertices in the ordered phases, $\chi_{tt}^{\rm R}(q,\omega)$ contributes directly to the self-energy; see also Fig.\ \ref{Fig1}(b) on this point.
Hence, the divergence of ${\rm Im}\chi_{tt}^{\rm R}(q,\omega)$ at $\omega=E_q$
implies that the lifetime of the quasiparticles with the dispersion relation $\omega=E_q$ is zero right on the dispersion curve, 
in contradiction to the original assumption of Eq.\ (\ref{rho_t-QP}).
This fact implies that the long-range transverse correlations in the continuous-symmetry-breaking phases 
generally prevent any gapless excitation from having a well-defined dispersion relation with an integer exponent 
below three dimensions at finite temperatures. 
There can be exceptional cases where well-defined dispersion relations are protected by some conservation laws.
Otherwise, cancellation of the infrared singularities to recover a sharp dispersion relation is unlikely to happen; 
no such mechanism at finite temperatures has been known to date.

It should be noted that the first and second terms in the curly brackets of Eq.\ (\ref{ImChi_tt-2}),  known as the Landau and Beliaev dampings,
respectively,  have certainly been investigated theoretically.\cite{HM65,Pitaevskii97,Giorgini98}
However, the damping rate in those finite-temperature studies is defined by a sum of Eq.\ (\ref{ImChi_tt-2}) over ${\bm q}$ for a given $\omega$ with no spectral resolution in terms of ${\bm q}$. Hence, the result obtained above has been overlooked.
Whereas the results of Ref.\ \onlinecite{HM65,Pitaevskii97,Giorgini98} show the stability of the entire system 
against the thermal activation of quasiparticle excitations, 
the present one reveals the presence of an intrinsic lifetime in each excitation which is not describable perturbatively.
Combined with the vanishing of the longitudinal mass, it tells us that 
the dispersion relation of the NG bosons may also be modified qualitatively at long wavelengths from that of the mean-field analysis.

\section{Detailed consideration on BEC}

In this section, we study in more detail how the $k^{-2}$ correlations modify
the spectral function of Eq.\ (\ref{rho_t-QP}) based on a formalism where the longitudinal susceptibility vanishes manifestly.
Specifically, we consider an isotropic homogenous Bose-Einstein condensate composed of identical particles with mass $m$ and spin $0$ at $T>0$ interacting via a two-body potential $U({\bm r})$,
which can be described in terms of the scalar fields $(\psi,\psi^*)\!\equiv\!(\psi_1,\psi_2)$.

\subsection{Vanishing of transverse and longitudinal masses}

Let us introduce Green's functions by $G_{jj'}(\vec{r},\vec{r}^{\,\prime})\!\equiv\!-\langle T_\tau \psi_j(\vec{r})\psi_{j'}(\vec{r}^{\,\prime})\rangle
+\langle \psi_j(\vec{r})\rangle\langle\psi_{j'}(\vec{r}^{\,\prime})\rangle$
with $j=1,2$ and $\vec{r}\!\equiv\!({\bm r},\tau)$, where $\tau$ lies in $[0,T^{-1}]$ with the periodic boundary conditions.\cite{AGD63,FW71}
The Fourier transform of $\hat G\!\equiv\! (G_{jj'})$ can be parametrized in units of $m\!=\!1/2$ as
\begin{align}
\hat G(\vec{k}) \!=\! \begin{bmatrix} -\Delta(\vec{k}) &  -i\varepsilon_n\!-\! k^2\!-\!\Sigma(-\vec{k})\!+\!\mu \\ i\varepsilon_n\!-\! k^2\!-\!\Sigma(\vec{k})\!+\!\mu & -\Delta(\vec{k}) 
\end{bmatrix}^{-1}
\label{hatG}
\end{align}
where $\mu$ is the chemical potential and $\vec{k}\equiv ({\bm k},i\varepsilon_n)$ in this expression.\cite{Kita19} 
The anomalous self-energy satisfies $\Delta(-\vec{k})\!=\!\Delta(\vec{k})$, and  the Hugenholtz-Pines relation\cite{HP59}
$\Sigma(\vec{0})-\mu=\Delta(\vec{0})$  ensures existence of a gapless excitation branch in the system.
Indeed,
choosing the condensate wave function $\Psi_j\!\equiv\!\langle\psi_j\rangle$ as real ($\Psi_1\!=\!\Psi_2\!\equiv\!\Psi$) and unitary-transforming
\begin{align}
\psi_j\!=\!\frac{1}{\sqrt{2}}\{\phi_l+i(-1)^{j-1}\phi_t\},
\label{Trans-lt}
\end{align}
we can decompose the field $\psi_j$ into the longitudinal and transverse ones $(\phi_l,\phi_t)$
with $\langle\phi_l\rangle\!=\!\sqrt{2}\Psi$ and  $\langle\phi_t\rangle\!=\!0$.
Elements of the transformed inverse Green's function are given by 
\begin{subequations}
\begin{align}
G^{-1}_{ll}(\vec{k})=&\,-\xi_{ll}(\vec{k}),
\\
G^{-1}_{tt}(\vec{k})=&\,-\xi_{tt}(\vec{k}),
\\
G^{-1}_{ lt}(\vec{k})=&\,-G^{-1}_{ tl}(\vec{k})=i\left\{i\varepsilon_n -\frac{\Sigma(\vec{k})-\Sigma(-\vec{k})}{2}\right\},
\end{align}
\end{subequations}
with 
\begin{align}
\begin{array}{l}\xi_{ll}(\vec{k}) \\ \xi_{tt}(\vec{k})\end{array}\biggr\}\equiv k^2+\frac{\Sigma(\vec{k})+\Sigma(-\vec{k})}{2}-\mu\pm \Delta(\vec{k}),
\end{align}
so that $\xi_{ll}(\vec{0})=2\Delta(\vec{0})$ and $\xi_{tt}(\vec{0})=0$ according to the Hugenholtz-Pines relation.
Thus, the transverse branch is gapless
and should behave as $\xi_{tt}({\bm k},0)\propto k^2$ for $k\rightarrow 0$.\cite{Hohenberg67}
The transformation (\ref{Trans-lt}) also clarifies that the Bogoliubov mode with the linear dispersion relation,
which is non-analytic in $k^2={\bm k}\cdot{\bm k}$,
is realized by the mixing of the gapless transverse branch $\xi_{tt}\propto k^2$ with the apparently gapped longitudinal one 
$\xi_{ll}(\vec{0})\propto 2\Delta(\vec{0})$ through a finite frequency $i\varepsilon_n$.
However, the longitudinal branch also becomes gapless as $\Delta({\bm k},0)\propto k^{d-4}$ for ${\bm k}\rightarrow {\bm 0}$
due to the transverse correlations.\cite{ABDS99,NN75}

Both the Hugenholtz-Pines relation and vanishing of the longitudinal {\it mass} $\Delta(\vec{0})=0$ can be derived based on Goldstone's theorem.\cite{GSW62,Weinberg95}
Specifically, the grand potential (or {\it effective action}) $\Gamma$ as a functional of the condensate wave function
is invariant under the global gauge transformation 
\begin{align*}
\delta\Psi_j(\vec{r})\!\equiv \! i\,\delta \theta\sum_{j'}(\hat\sigma_3)_{jj'} \Psi_{j'}(\vec{r})
\end{align*}
by the phase angle $\delta\theta$, where
$\hat\sigma_3$ denotes the third Pauli matrix and the $\vec{r}$ dependence originates from a coupling of $\psi_j$ to an external source field $J_j(\vec{r})$.\cite{Weinberg95,Amit84,ABDS99}
This invariance of $\Gamma$ reads 
\begin{align*}
\sum_{jj'}\int d^4 r \frac{\delta\Gamma}{\delta\Psi_{j'}(\vec{r})}(\hat\sigma_3)_{j'j}\Psi_{j}(\vec{r})\!=\! 0.
\end{align*}
Differentiating the equality $n$ times in terms of $\Psi_{j_i^{}}(\vec{r}_{j_i^{}})$ ($i\!=\!1,2,\cdots,n$), switching off the source field, and adopting the $\vec{k}$ representation, we obtain \cite{Amit84,ABDS99,Kita19}
\begin{align}
&\,\sum_{i=1}^n \sum_{j_i'}\Gamma^{(n)}_{j_1\cdots j_i'\cdots  j_n}(\vec{k}_1,\cdots,\vec{k}_n)(\hat\sigma_3)_{j_i'j_i^{}}
\notag \\
&\,+\sum_{j'j}\Gamma^{(n+1)}_{j_1\cdots j_n j'}(\vec{k}_1,\cdots,\vec{k}_n,\vec{0})(\hat\sigma_3)_{j'j}\Psi_{j}=0,
\label{Gamma-identity}
\end{align}
where $\Gamma^{(n)}_{j_1\cdots j_n}(\vec{k}_1,\cdots,\vec{k}_n)\!\propto\! \delta_{\vec{k}_1+\cdots+\vec{k}_n,\vec{0}}$ is the $n$-point vertex defined by the Fourier transform of 
$\delta^n\Gamma/\delta\Psi_{j_1}(\vec{r}_1)\cdots\delta\Psi_{j_n}(\vec{r}_n)$. 
The $n\!=\!1$ case of Eq.\ (\ref{Gamma-identity}) is the identity known standardly as Goldstone's theorem \cite{Weinberg95} and
also equivalent to the Hugenholtz-Pines relation.\cite{HP59} The latter fact can be seen by substituting 
the stationarity condition
$\Gamma^{(1)}_j(\vec{0})=0$ of the grand potential and also the connection $\hat\Gamma^{(2)}(-\vec{k},\vec{k})=-\hat{G}^{-1}(\vec{k})$ between
$\hat\Gamma^{(2)}\equiv (\Gamma^{(2)}_{jj'})$ and Eq.\ (\ref{hatG}).\cite{Amit84,Weinberg95}
On the other hand, the case of $n=2$ in Eq.\ (\ref{Gamma-identity}) yields the vanishing of $\Delta(\vec{0})$.
To see this, it is convenient to transform Eq.\ (\ref{Gamma-identity}) into  the $(\phi_l,\phi_t)$ representation through $\hat\sigma_3\rightarrow \hat\sigma_2$ and $\Psi_j\rightarrow \delta_{jl}\sqrt{2} \Psi$,
where the Hugenholtz-Pines relation is given by
\begin{align}
\Gamma^{(2)}_{tt}(\vec{0},\vec{0})=0. 
\label{HP}
\end{align}
The choice $(j_1,j_2)=(l,t)$ then yields
\begin{align*}
\Gamma^{(2)}_{ll}(-\vec{k},\vec{k})-\Gamma^{(2)}_{tt}(-\vec{k},\vec{k})=\sqrt{2}\Gamma^{(3)}_{ltt}(-\vec{k},\vec{k},\vec{0})\Psi
\end{align*}
so that 
\begin{align}
\Gamma^{(2)}_{ll}(\vec{0},\vec{0})=\sqrt{2}\Gamma^{(3)}_{ltt}(\vec{0},\vec{0},\vec{0})\Psi
\end{align}
holds according to Eq.\ (\ref{HP}).
On the other hand, the self-energies can be expressed graphically as Fig.\ 1.\cite{Kita21}
The graph (b) contains the transverse contribution $c_1 \chi_{tt}(\vec{0})\Gamma^{(3)}_{ltt}(\vec{0},\vec{0},\vec{0})$ to  $\Gamma^{(2)}_{ll}(\vec{0},\vec{0})$
given in terms of some constant $c_1$, 
whereas the other processes yield a finite value $c_0$.
Hence, $\Delta(\vec{0})=\Gamma^{(2)}_{ll}(\vec{0},\vec{0})/2$ can be written as 
\begin{align}
\Delta(\vec{0})=\frac{c_0/2}{1-c_1 \chi_{tt}(\vec{0})/\sqrt{2}\Psi},
\end{align}
where $\chi_{tt}(\vec{0})$ diverges for $d<4$ at finite temperatures as seen by inspecting Eq.\ (\ref{chi_tt}) with $G_{tt}(k,0)\propto k^{-2}$ for $k\rightarrow 0$.
Thus, we conclude that $\Delta(\vec{0})=0$ holds for $d<4$ at $T>0$.\cite{Kita19}
This argument is the finite-temperature extension of the one given by Nepomnyashchi\u{i} and Nepomnyashchi\u{i} \cite{NN75}  
that found $\Delta(\vec{0})=0$ for $d\leq 3$ at $T=0$.

\subsection{Numerical study}

The qualitative argument given above clarifies the crucial importance of the identities of Eq.\ (\ref{Gamma-identity})
for studying the excitations of BECs.
We now perform a more detailed calculation of its spectral function 
by adopting a contact potential $U({\bm r})=U\delta({\bm r})$.
The formalism developed previously \cite{Kita21} so as to satisfy Eq.\ (\ref{Gamma-identity})
yields the following expressions for the self-energies in the weak-coupling regime
\begin{align}
\Sigma(\vec{k})-\mu= \Delta(\vec{k})=\Psi^2 \tilde{U}(\vec{k})
\label{SE}
\end{align}
given in terms of the effective potential
\begin{align}
\tilde{U}(\vec{k})=\frac{U}{1-U[2\chi_{GG}^{}(\vec{k})\!+\!2\chi_{FF}^{}(\vec{k})\!-\!\chi_{GF}^{}(\vec{k})\!-\!\chi_{FG}^{}(\vec{k})]},
\label{tU}
\end{align}
where $\chi_{GF}^{}(\vec{k})$, for example, is obtained from Eq.\ (\ref{chi_tt}) by the replacement
$(G_t,G_t)\rightarrow (G,F) \equiv (G_{12}, -G_{11})$ given in terms of the upper elements of Eq.\ (\ref{hatG}).
Equation (\ref{tU}) is  compatible with all the qualitative results mentioned in the previous paragraph, as seen by noting $G(k,0)\approx F(k,0)\approx \frac{1}{2}G_{tt}(k,0)$ for $k\rightarrow 0$.
The Bogoliubov theory is reproduced from Eq.\ (\ref{SE}) by omitting the correlation effects as $\tilde{U}(\vec{k})\rightarrow U$.
Indeed, substituting the corresponding self-energy into Eq.\ (\ref{hatG}), we obtain the Bogoliubov spectrum 
\begin{align}
E_{\rm B}(k)=k\sqrt{k^2+2\Delta_{\rm B}},\hspace{10mm}\Delta_{\rm B}\equiv \Psi^2 U,
\label{E_B}
\end{align}
which has a linear dispersion relation for $k\xi \!\lesssim \! 0.5$ where $\xi\equiv \Delta_{\rm B}^{-\frac{1}{2}}$.
Below we present results of the first iteration based on 
Eqs.\ (\ref{hatG}), (\ref{SE}) and (\ref{tU}) starting from the Bogoliubov theory, 
in which $\chi_{GG}^{}(\vec{k})$ etc.\ are estimated by the one-loop approximation given by Eq.\ (\ref{chi_tt}).
We have adopted a weak-coupling interaction of $\bar{n}U/T_0=0.1$ ($\bar{n}$: particle density),
where we can also approximate $\Psi^2\approx [1-(T/T_0)^{3/2}]\bar{n}$ using
the non-interacting condensation temperature $T_0$.

\begin{figure}
\begin{center}
\includegraphics[width=0.8\linewidth]{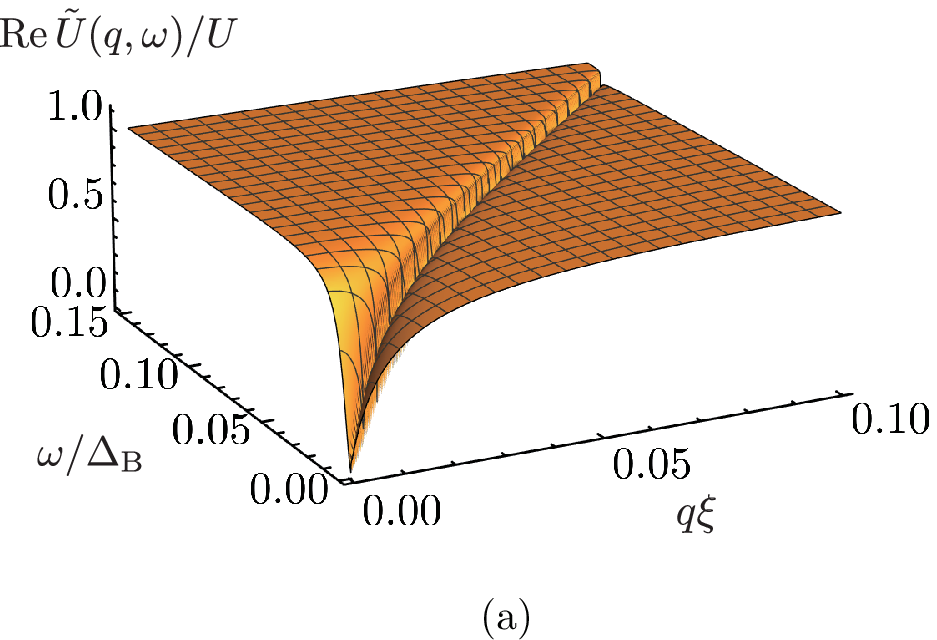}
\includegraphics[width=0.8\linewidth]{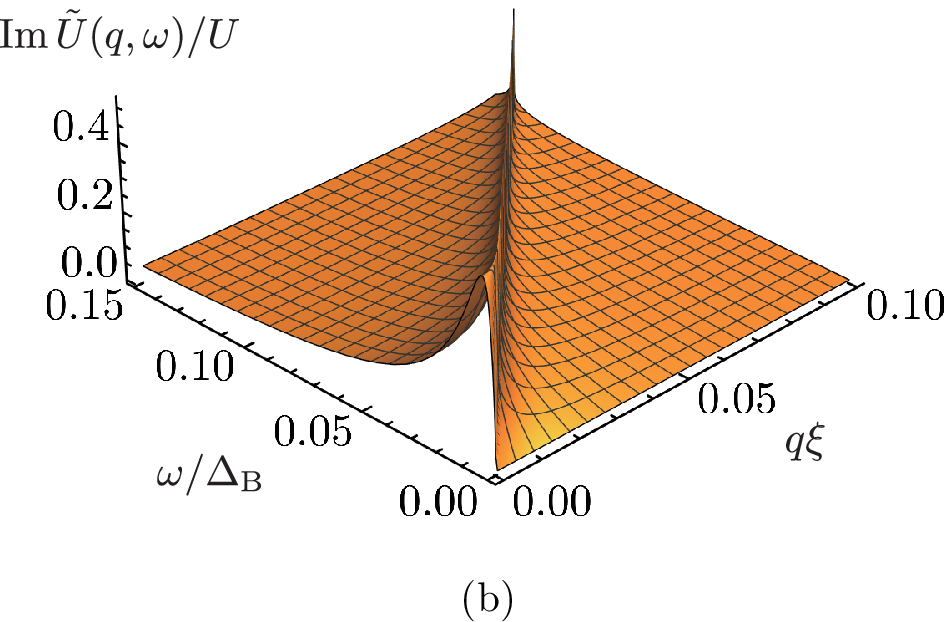}

\vspace{5mm}
\includegraphics[width=0.7\linewidth]{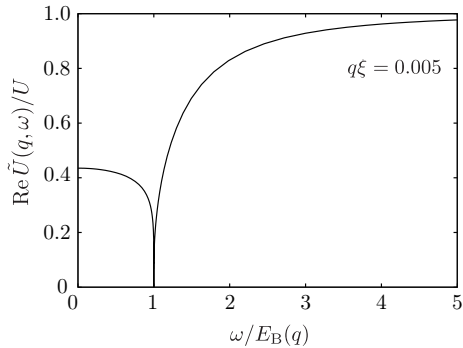}

\vspace{5mm}
\includegraphics[width=0.7\linewidth]{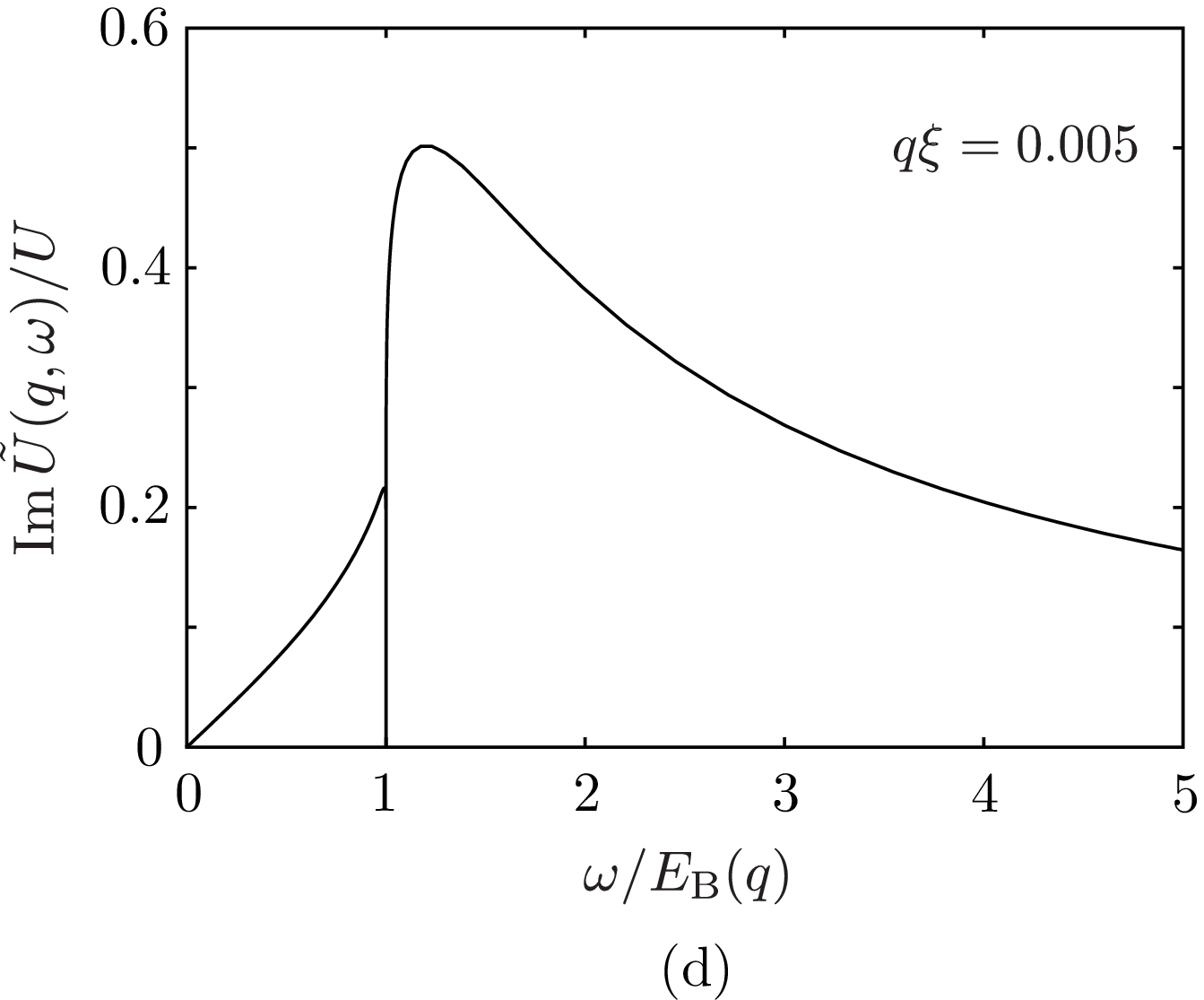}
\end{center}
\caption{(Color online) Normalized effective potential $\tilde{U}(q,\omega)/U$ at $T/T_0=0.3$.
Three-dimensional plots of (a) the real part and (b) the imaginary part. 
Detailed plots of (c) the real part and (d) the imaginary part as functions of  $\omega/E_{\rm B}(q)$ at $q\xi=0.005$.}
\label{Fig2}
\end{figure}

\begin{figure}[t]
\begin{center}
\includegraphics[width=0.8\linewidth]{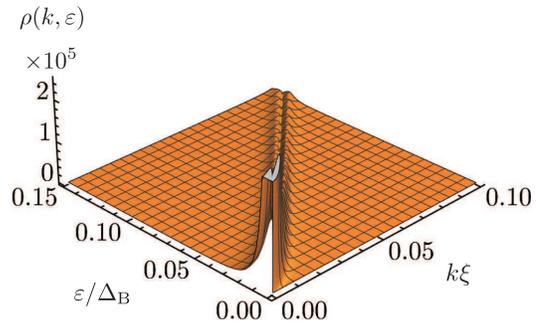}

\vspace{5mm}
\includegraphics[width=0.7\linewidth]{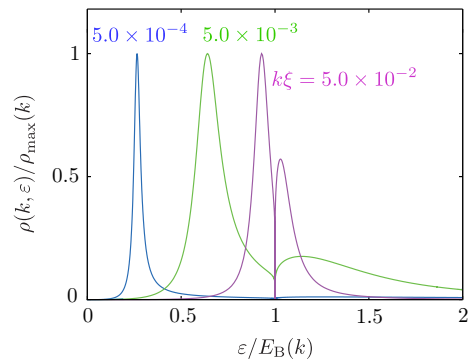}
\end{center}
\caption{(Color online) (a) Spectral function $\rho(k,\varepsilon)\!\equiv \! -2{\rm Im}G^{\rm R}(k,\varepsilon)$ at $T/T_0=0.3$. 
(b) Detailed plots of the normalized spectral function as a function of $\varepsilon/E_{\rm B}(k)$ for three different values of $k\xi=5.0\times 10^{-2}$, $5.0\times 10^{-3}$, $5.0\times 10^{-4}$.}
\label{Fig3}
\end{figure}

Figure \ref{Fig2} shows the normalized effective potential $\tilde{U}(q,\omega)/U$ 
calculated at $T/T_0=0.3$.
In Fig.\ \ref{Fig2}(a), a vertical fault with a deep steep valley is seen to develop in the plateau of ${\rm Re}\,\tilde{U}$ along the Bogoliubov dispersion $\omega\!=\! E_{\rm B}(q)$.
The imaginary part ${\rm Im}\,\tilde{U}$ of Fig.\ \ref{Fig2}(b), on the other hand, is characterized by sharp mountainous peaks along $\omega\!=\! E_{\rm B}(q)$ that rise and broaden increasingly towards the origin. 
A closer look at the region of $\omega\!=\!E_{\rm B}(q)$ reveals that both ${\rm Re}\,\tilde{U}$ and  ${\rm Im}\,\tilde{U}$ vanish along the Bogoliubov dispersion due to the divergence of the denominator in Eq.\ (\ref{tU}), as seen in Fig.\ 2(c) and (d) calculated for $q\xi=0.005$.
The approaches to $0$ are logarithmic and non-analytic in three dimensions.
Thus, the first iteration predicts that the vanishing of the longitudinal mass $\Delta(0,0)\!=\!0$ extends to the finite $(q,\omega)$ region along the Bogoliubov dispersion
curve.
Figure \ref{Fig3} shows the corresponding spectral function $\rho(k,\varepsilon)\equiv -2{\rm Im}G^{\rm R}(k,\varepsilon)$.
The spectral weight vanishes along the Bogoliubov dispersion
to produce a double-peak structure for each $k$ as a function of $\varepsilon$.
Moreover, the higher peak shifts towards zero frequency as $k\rightarrow 0$
due to $\tilde{U}(0,0)=0$. 
The fault in Fig.\ \ref{Fig2}(a) decreases in magnitude as $T\rightarrow 0$ 
to disappear completely at zero temperature, 
but $\tilde{U}(0,0)$ still vanishes logarithmically (i.e., non-analytically) even in the limit.\cite{NN75}

The above results of the first iteration elucidate essential modifications to the mean-field dispersion relation by the transverse correlations.
Most notably, the spectral weight has disappeared completely along the mean-field dispersion curve with the $\delta$-function spectrum.
They indicate presence of an intrinsic lifetime in the excitations
and a qualitative change in the dispersion relation from the mean-field analytic one at long wavelengths
due to the vanishing of the longitudinal mass, 
as seen by setting $\Delta_{\rm B}\rightarrow 0$ for $k\rightarrow 0$  in Eq.\ (\ref{E_B}).
Note that the longitudinal mass vanishes even at zero temperature logarithmically, so that the latter statement holds true even at zero temperature. An intrinsic lifetime in the excitations may also persist down to $T\!=\! 0$, as for the case of some sorts of magnon systems.\cite{CZ09}

Experiments on the Bogoliubov excitations of Eq.\ (\ref{E_B}) have been performed for trapped Bose-Einstein condensates.\cite{Vogels02,Ozeri02,OKSD05}
The particle-hole mixing of Eq.\ (\ref{chi_tt}) predicted by the Bogoliubov theory has been observed clearly at $k\xi=0.38$,\cite{Vogels02,OKSD05} 
and a reasonable agreement between the Bogoliubov theory and measured excitation energies has been reported for $k\xi\gtrsim 0.5$.\cite{Ozeri02,OKSD05}
However, the observed excitation spectra are accompanied by substantial broadenings of the order of the 
excitation energies themselves,\cite{Vogels02,OKSD05} which may be caused partly by the inhomogeneity of the trap potential.
Hence, they cannot tell whether the dispersion relation is really a sharp distinct one or accompanied by some essential broadening
and deviation from Eq.\ (\ref{E_B}), especially at the low excitation energies of $k\xi\lesssim 0.1$.
It remains to be performed both theoretically and experimentally to develop methods to detect them in high resolutions 
and clarify the excitation spectrum definitely.

\section{Concluding Remarks}

We have studied how the long-range transverse correlations affect the dispersion relation of Nambu-Goldstone bosons at finite temperatures
by incorporating the correlations perturbatively at the one-loop level.
It is found that the divergence of the longitudinal susceptibility (i.e., vanishing of the longitudinal {\em mass}), which is
characteristic of systems with broken continuous symmetries, extends to the finite energy-momentum region
as vanishing of the spectral weight right on the assumed dispersion curve below three dimensions at finite temperatures.

We finally point out the necessity of distinguishing various correlation functions 
that have been focused in respective studies of Goldstone's theorem.
Specifically, its first proof,\cite{GSW62,Weinberg95} given around Eq.\ (\ref{Gamma-identity}), is relevant 
to the two-point  function of the field $\psi_j(\vec{r})$, whereas the second proof \cite{GSW62,Weinberg95}
concerns the correlations of $\psi_j(\vec{r})$ with the field $j_\alpha^0(\vec{r})$ that constitutes the conserved charge $Q_\alpha \equiv \int  j_\alpha^0(\vec{r})\,d^3 r$. Finally, the dispersion relations of the NG bosons have been discussed
in terms of the two-point functions of $j_\alpha^0(\vec{r})$.\cite{NC76,Schafer01,WB11,WM12,Hidaka13} 
The three kinds of correlation functions are generally different and can have different poles.
This may be realized clearly in terms of the single-component BECs where $j_\alpha^0(\vec{r})$ is the density operator $n(\vec{r})=\psi_2(\vec{r})\psi_1(\vec{r})$.
The fact that its four-point vertices have different $q$- and $\omega$-limits,\cite{Kita21}
contrary to the assumption made by Gavoret and Nozi\`eres,\cite{GN64} implies that 
the system can embrace collective modes in the correlation functions of $(n,\psi)$ and $(n,n)$ that are distinct from the single-particle excitations in $(\psi,\psi)$, similarly as in the case of Fermi liquids and superfluid Fermi liquids.\cite{AGD63,SR83}
The topic here is relevant to the poles of  $(\psi,\psi)$ concerning the first proof.

\section*{Acknowledgment}

This work was supported by JSPS KAKENHI Grant Number JP20K03848.

\end{document}